\documentclass[aps,prl,showpacs,amssymb,superscriptaddress,footinbib,twocolumn]{revtex4-1}
\usepackage{graphicx}
\usepackage{amsmath}
\usepackage{color}
\usepackage{cancel}
\usepackage{natbib}
\usepackage{ulem} 
\usepackage{tikz}
\newcommand{\ket}[1]{\ensuremath{\left|{#1}\right\rangle}}
\newcommand{\bra}[1]{\ensuremath{\left\langle{#1}\right|}}
\newcommand{\braket}[2]{\ensuremath{\langle{#1}|{#2}\rangle}}

\usepackage{mathptmx} 

\begin{document}

\title{Exploiting structured environments for efficient energy transfer: \\ The phonon antenna mechanism}

\author{Marco del Rey}
\affiliation{Instituto de F\'{i}sica Fundamental, CSIC, Serrano 113-B, 28006 Madrid, Spain}

\author{Alex W. Chin}
\affiliation{Theory of Condensed Matter Group, University of Cambridge, J.~J. Thomson Avenue, Cambridge, CB3 0HE, United Kingdom}

\author{Susana F. Huelga}
\affiliation{Institut f\"{u}r Theoretische Physik, Albert-Einstein Allee 11, Universit\"{a}t Ulm, 89069 Ulm, Germany}
\affiliation{Center for Integrated Quantum Science and Technology, Albert-Einstein Allee 11, Universit\"{a}t Ulm, 89069 Ulm, Germany}

\author{Martin B. Plenio}
\affiliation{Institut f\"{u}r Theoretische Physik, Albert-Einstein Allee 11, Universit\"{a}t Ulm, 89069 Ulm, Germany}
\affiliation{Center for Integrated Quantum Science and Technology, Albert-Einstein Allee 11, Universit\"{a}t Ulm, 89069 Ulm, Germany}

\date{\today}

\begin{abstract}
A non-trivial interplay between quantum coherence and dissipative environment-driven dynamics is becoming increasingly recognised as key for efficient energy transport in photosynthetic pigment-protein complexes, and converting these biologically-inspired insights into a set of design principles that can be implemented in artificial light-harvesting systems has become an active research field.  Here we identify a specific design principle -\textit{ the phonon antenna} -  that demonstrates how inter-pigment coherence is able to modify and optimize the way that excitations spectrally sample their local environmental fluctuations. We place this principle into a broader context and furthermore we provide evidence that the Fenna-Matthews-Olson complex of green sulphur bacteria has an excitonic structure that is close to such an optimal operating point, and suggest that this general design principle might well be exploited in other biomolecular systems.

\end{abstract}
\maketitle

\textit{Introduction.---} Experimental techniques such as optical 2D Fourier Transform
spectroscopy have recently begun to probe the ultrafast photophysics of energy transport
in a range of pigment-protein complexes (PPCs) taken from photosynthetic green sulphur
bacteria, marine algae and higher plants \cite{gregory2007evidence,panitchayangkoon2010long,calhoun2009quantum,collini2010coherently}.
All of the key photosynthetic light reactions (photon capture, energy transport and charge
generation) are carried out in PPC structures \cite{blankenship2002molecular,van2000photosynthetic},
and the often $>90\%$ quantum efficiency with which they carry out these functions has created
considerable interest in understanding and exploiting their design features for artifical
solar energy applications \cite{scholes2011lessons,blankenship2002molecular}.
Surprisingly, recent experiments have found direct evidence for long-lasting quantum
coherence amongst the excitons which transport energy in these complexes. In  the case of
the Fenna-Matthews-Olson (FMO) complex, these coherences can persist on picosecond timescales
in cryogenic conditions and are still observable at room temperatures \cite{panitchayangkoon2010long,hayes2010dynamics}.
Several phenomenological theories have subsequently shown that there is an optimal mixture
of coherent inter-pigment energy transport and stochastic environmental noise that may lead
to faster and higher-yield energy delivery in PPC architectures, suggesting that quantum
effects may underpin their efficient function \cite{nat2, plenio08, caruso09, ishizaki2010quantum, olaya2008efficiency, chindc2010, carusocd10}. Microscopic investigations have also recently shown the key role of both discrete and
continuous environmental fluctuation spectra in stabilising the long-lasting coherences
observed in spectroscopy as well as facilitating efficient transport
\cite{chin2012vibrational, christensson2012origin, kramer12, chinproc12, kolli2012fundamental}.

In this letter we build upon these microscopic studies to describe an explicit design
principle through which coherent excitonic coupling is \textit{exploited} to optimize
energy transport in the typical environmental conditions found in PPCs. This is based
on the idea that coherent couplings can allow a pigment network to be spectrally `tuned'
into configurations where the emerging relaxation (transport) pathways may extract maximum
noise strength from the environment and thus proceed faster. In analogy to the well-established
idea that excitonic interactions in light-harvesting antenna complexes allow pigments to
absorb a broader range of the ambient light spectrum \cite{blankenship2002molecular,van2000photosynthetic},
we refer to this principle of advantageous, coherently-modified sampling of the protein noise
spectrum as \textit{the phonon antenna mechanism} \cite{chinproc12}.

\textit{The phonon antenna.---}
\begin{figure}
\includegraphics[width=8cm]{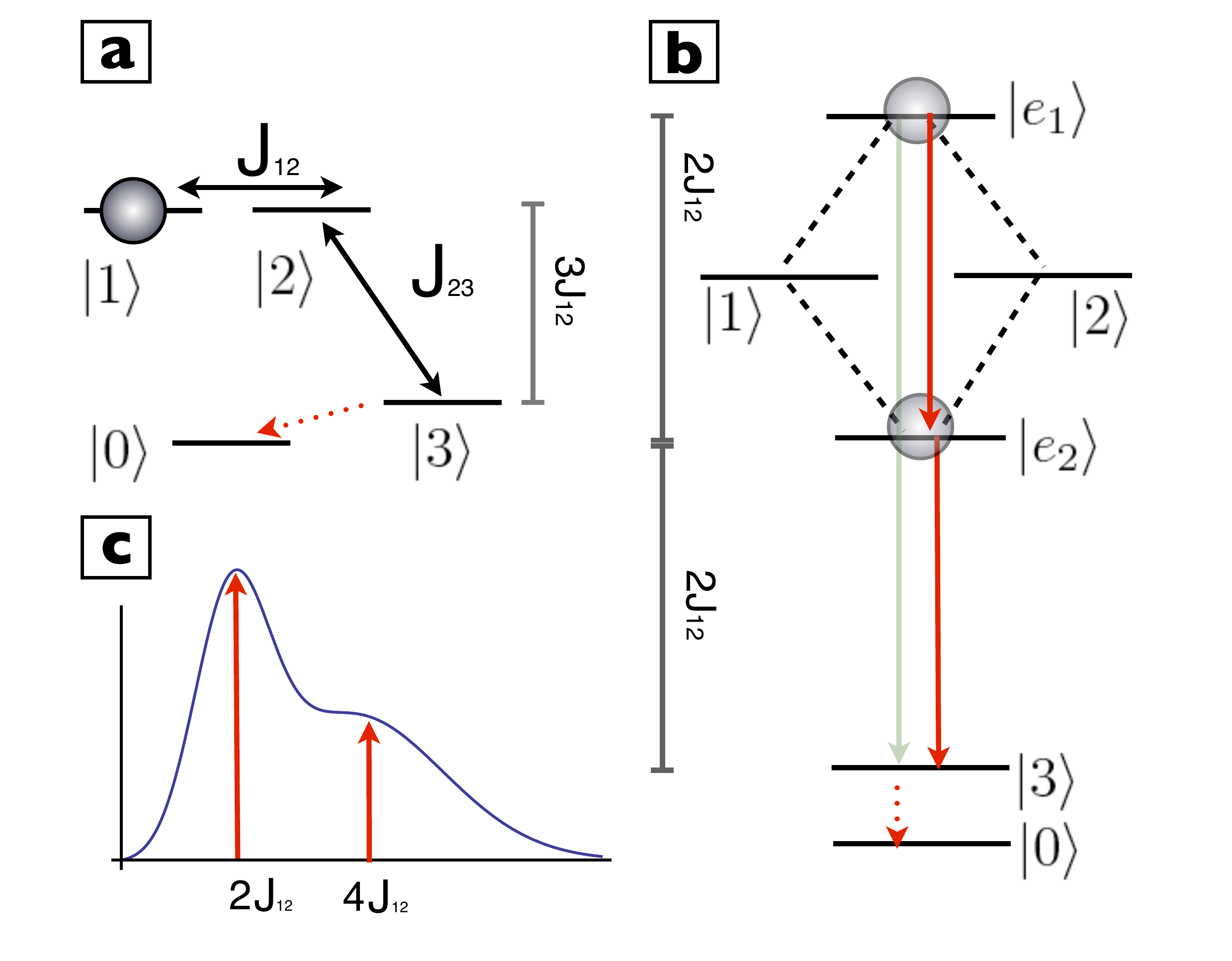}
\caption{(a) A model system consisting of a three-site network showing coherent couplings
(double-headed arrows) and relative local site energies of state $|i\rangle$ subject to
dephasing noise as described
in the text. An excitation on site $1$ will be transferred to the sink (site $0$) via sites
$2$ and $3$. The population transfer from $3$ to the sink is incoherent (dotted arrow). (b)
Diagonalising the network Hamiltonian results in new eigenstates $|e_{i}\rangle$. For the
parameters in (b), the energy differences between eigenstates correspond to maxima of a
hypothetical environmental spectral function shown in (c). Environment-induced relaxation
between these states (single-headed arrows) is thus optimised, as transition rates are
proportional to the spectral density at the exciton energy differences. }
\label{antennapic}
\end{figure}
To illustrate the concept, we begin by setting out the (canonical) model of exciton transport
in a PPC network whose function is to deliver excitonic energy to a reaction centre (where it
is used to generate electrons for subsequent  -- dark -- stages of the photosynthesis)
\cite{van2000photosynthetic}. The network consists of a set of chromophores (sites) which are
held in space by a protein matrix. Each chromophore possesses an optically excited state $\ket{i}$,
whose energy $\epsilon_{i}$ depends on the local environment. The electronic Hamiltonian
$H_S$ is then given by $H_S = \sum_i \epsilon_i \ket{i}\bra{i} + \sum_{i\neq j } J_{ij} \ket{i}\bra{j}$,
the sum of single exciton contributions on each site $i$ with local energy energy $\epsilon_{i}$ and
coherent coupling terms describing the inter-site exchange of excitations with dipolar coupling
strength $J_{ij}$. The states obeying $H_S\ket{e_n} = E_n \ket{e_n}$ are in general delocalized
excitonic states, which can be also be expressed in the site basis as $\ket{e_n}=\sum_{i}C^{i}_{n}|i\rangle$.
The ultimate transfer of the exciton to a reaction center is modelled by including an isolated site
(labeled $0$ and referred to as the `sink') which is populated irreversibly from a particular local
site \cite{plenio08,nat2, caruso09}.

Non-electronic degrees of freedom, due to the surrounding protein environment and/or internal vibrational
modes of the chromophores, also couple to the excitons, inducing local fluctuations of the site energies
$\epsilon_{i}$. These `environmental' fluctuations (phonons) are modelled as originating from a continuous
\textit{bath} of harmonic oscillators coupled to the exciton and constitute the primary source of energy
relaxation and dephasing for the excitonic system \cite{van2000photosynthetic, adolphsrenger06} - see
Appendix A. Coherent inter-site couplings in $H_S$ can be stronger or of the same order than the characteristic
strength of the environmental interactions \cite{brixner2005two, gregory2007evidence, calhoun2009quantum, collini2010coherently}.
In such cases delocalised exciton eigenstates, for which $H_S$ is diagonal, therefore provide the \textit{preferred}
basis to analyze the system's transport dynamics. Within this framework, we can employ secular Redfield theory
to obtain the following Markovian equation of motion for the population $\rho_{nn}$ of the exciton states
$\{ \ket{e_n}\}$ \cite{redfield, blum}:
\begin{align}
\dot{\rho}_{mm}\! &= \! \sum_{n \neq m} \! W_{mn}\rho_{nn} \!  - \! \sum_{n \neq m}\!  (W_{nm}\! +\! \Gamma_{e_m 0}) \rho_{mm}, \quad m \neq 0 \nonumber\\
\dot{\rho}_{00} &= \sum_{n \neq 0} \Gamma_{e_n \rightarrow 0} \rho_{nn} ,
\end{align}
where the transition rates between eigenstates are given by:
\begin{align}
W_{mn} &= \Big\{\begin{array}{ll}
          2\pi  \mathcal J(\omega_{mn})\chi_{mn} n(\omega_{mn}),  & \mbox{$\omega_{mn} \geq 0$}\label{rate}\\
         2\pi  \mathcal J(\omega_{nm})\chi_{mn} [n(\omega_{nm})+1], & \mbox{$\omega_{mn} < 0$}\end{array}
\end{align}
with
\begin{align}
\chi_{mn} & =\sum_i |C^i_n C^i_m|^2 ,\  n({\omega}) =\frac{1}{e^{\frac{\omega}{k_B T} -1 }}, \ \omega_{mn} = E_m - E_n.\nonumber
\end{align}
Here the irreversible process that locally populates the sink from site $n$ is assumed to have an (environment-independent)
rate $\Gamma_{n \rightarrow 0} = 1/{(1 \ ps)}$ \cite{plenio08,caruso09,nat2}, which is typical for transfer to reaction
centers. In the exciton picture this leads to population transfer to the sink at a rate $\Gamma_{e_n \rightarrow 0}
= |\braket{e_n}{n}|^2\Gamma_{n \rightarrow 0} $.
Our figure of merit for assessing the transport efficiency of a given network is the population in the sink
$p_{\text{sink}}(t)=\rho_{00}(t)$ at some time $t$ following the local excitation of a selected site. The
population dynamics is entirely driven by the environment and, as a result of energy conservation, the relaxation
rate of an exciton transition $|e_{n}\rangle\rightarrow |e_{m}\rangle$ (which requires the emission of a phonon)
is thus dependent on both the phonon density of states and the coupling strength to phonons at frequencies equal
to the energy difference $E_{nm}$. Both these properties are described by the environmental spectral function
$\mathcal{J}(\omega)$ and the transition rate $W_{nm}$ is proportional to $\mathcal{J}(E_{nm})$. As typical
PPC spectral functions often have (at least) one maximum at a finite frequency \cite{modes} and may exhibit
sharp features due to long-lived vibrational modes in the environment, it becomes possible to \textit{tune}
the Hamiltonian parameters to create excitonic states with energy differences $E_{nm}$ at local maxima of the
spectral function and thus optimise $W_{nm}$. {\it This is the basis of the phonon antenna effect}.

Figure \ref{antennapic} illustrates the phonon antenna mechanism in a sequential transport scenario. The simplest
model consists of a network made up of three sites, $|i\rangle$,  with local energies $\epsilon_i$, $i=1,2,3$, as
shown in Figure \ref{antennapic}a. Sites $1$ and $2$ are assumed to be nearly degenerate and strongly coupled, with
a smaller coupling between sites 2 and 3, and almost no coupling between sites 1 and 3; that is, $J_{13}=0, J_{23}
\ll J_{12} \leq \epsilon_1 \sim \epsilon_2=\epsilon, \epsilon_3=0$.  Initially, only site $|1\rangle$ is excited.
For these parameters, the coherent coupling $J_{12}$ splits the degeneracy of states $|1\rangle$ and $|2\rangle$
and the  exciton eigenstate of $H_S$ are approximately $|e_{1}\rangle\approx 2^{-1/2}(|1\rangle+|2\rangle)$,
$|e_{2}\rangle\approx 2^{-1/2}(|1\rangle-|2\rangle)$ ,$|e_{3}\rangle\approx |3\rangle$, with energies $E_1=
\epsilon+J_{12},E_{2}=\epsilon-J_{12}$ and $E_{3}=0$. The initial condition can be re-expressed as $|1\rangle=
2^{-1/2}(|e_1\rangle+|e_2\rangle)$, and an energy diagram for these new states and initial condition is shown
in Figure \ref{antennapic}b. The efficiency of transport to site $3$ (and thence to the sink) now depends on the
dissipative rates $W_{12}$ and $W_{23}$ of the environment-induced transitions $|e_1\rangle\rightarrow|e_2\rangle$
and $|e_2\rangle\rightarrow |3\rangle$, respectively.

Let us consider now an spectral function possessing a maximum at an energy $\omega_{H}$, as depicted in Fig.
\ref{antennapic}c). When $J_{12}=\epsilon/2=\omega_{H}$, we find that $E_{12}  \simeq E_{23} \simeq \omega_{H}$ and a ladder
of equally spaced states appears with a splitting $\omega_{H}$. Thus a clever choice of $J_{12}$ simultaneously
optimizes \textit{both} transition rates $W_{12}$ and $W_{23}$ and generates efficient energy transport to the sink.
This simple system illustrates the essential idea behind implementing an efficient phonon antenna: \textit{
Electronic interactions are tuned (e.g. by controlled or evolutionary adaption of distances and relative
orientation of chromophores) to achieve resonance with maxima of the environmental vibrational spectra}.

This already suggests how in more general networks, coherent couplings and site energies could be tuned to
enhance (or inhibit) a particular set of energetic transitions  (which in a multi-chromophore network could
also correspond to a particular spatial pathway). The more structure in the spectral function, the greater
the possibilities for phonon antenna effects to be harnessed, as shown in Fig.\ref{antennapic}c where a
shoulder feature in $\mathcal{J}(\omega)$ also enhances the direct $W_{13}$ transition. In anticipation of
our analysis of the FMO complex, we now perform a quantitative analysis of this three-site model system.

Motivated by actual values in typical PPCs, the considered Hamiltonian parameters are, in units of $cm^{-1}$:
$\epsilon_1=\epsilon_2=300, \ J_{12}=J_{21}=100, \ J_{23}=J_{32}=30, \ \epsilon_3=J_{13}=J_{31}=0.$ The spectral
density function is taken to be of a quasi-Lorentzian shape centered in $\omega_H$ \cite{garg}:
\begin{equation}
\mathcal J(\omega) = \frac{2\beta \omega \omega_H^4}{(\omega^2-	\omega_H^2)^2+(2 \pi \Gamma \omega)^2}
\end{equation}
where $\beta$ is fixed in order to have a reorganization energy $\lambda\simeq \frac{\pi\omega_H^2\beta}{\Gamma} =
35 \text{ cm}^{-1}$, and $\Gamma=60 \text{ cm}^{-1}$. We assume that initially site 1 is selectively excited.

\begin{figure}[h]
\begin{center}
\includegraphics[width=0.45\textwidth]{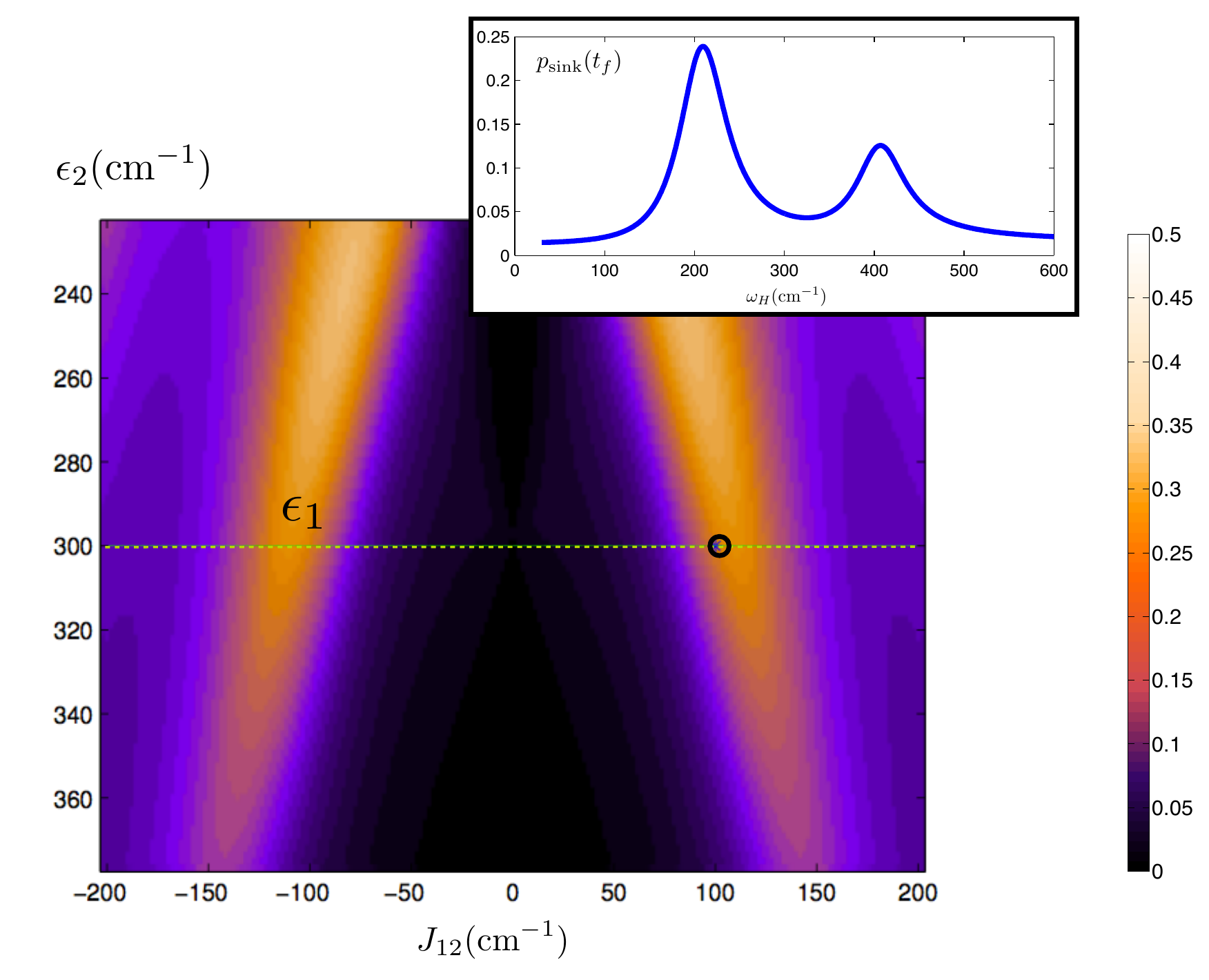}
\end{center}
\caption{Transferred sink population after 2 ps for the Markovian 3 site model system described in the text and depicted in Figure 1. The environment is described in terms of a Lorentzian spectral density peaked at a frequency $\omega_H=200 \text{cm}^{-1}$ for a temperature $T=4 K$. Actual values of the Hamiltonian values, lie as expected (blue dot), at the top of a maximum in transfer yield. In the inset, population in the sink after 2 ps for a spectral density peaked at $\omega_H$, given a temperature $T=4 K$. As an example were the matching takes place, the phonon antenna is formed and adjusted to a peak $\omega_H\simeq 2J = 200 \text{ cm}^{-1}$. The other peak around $\omega_H\simeq 400 \text{ cm}^{-1}$ but with half the height  corresponds to an antenna effect directly connecting sites $\ket{e_1}$ and $\ket{e_3}$. Small deviations come from $J_{23}$  being not zero. }
\label{fig:markov3sitescouplings}
\end{figure}

Within this parameter regime, the inset in Figure \ref{fig:markov3sitescouplings} shows the population in the
sink after $2$ ps as $\omega_{H}$ is varied. As expected, a clear peak in the transport efficiency emerges when
the spectral function is peaked slightly above the value  $\omega_{H}=200 \mathrm{cm}^{-1}$ which would correspond
to perfect matching of the inter-exciton energy differences to the maximum of the spectral density. Figure
\ref{fig:markov3sitescouplings} shows the sink populations after $2$ ps for fixed $\omega_{H}$ and $\epsilon_{1}$,
while varying the coupling $J_{12}$ and the site energy $\epsilon_{2}$.  We observe a symmetric structure in the
transport efficiency with respect to $J_{12}\rightarrow\ -J_{12}$, but find that the degenerate configuration
$\epsilon_1=\epsilon_2$ is not a global maximum of the efficiency. The highest efficiency occurs for
$\epsilon_2\approx240 \mathrm{cm}^{-1}$ and $J_{12}\approx95 \mathrm{cm}^{-1}$. This larger value of the
efficiency is the result of a slightly enhanced coupling (overlap) between sites $2$ and $3$ that results
from shifting site $2$ to a lower (local) energy. However, the differences in exciton energies $E_{1}-E_{2}=200
\mathrm{cm}^{-1}$ and $E_{2}-E_{3}=180 \mathrm{cm}^{-1}$ remain close to the maximum of the spectral density.
Indeed we notice that the reduction in the optimal values of $J_{12}$ as $\epsilon_{2}$ decreases maintains energy
differences close to the spectral density maximum. The shift in the global maximum illustrates the role of the
$\chi$ factor in determining the energy and coupling structures which optimally exploit the functional form of
the environment's spectral function.

\textit{The phonon antenna in the FMO complex: Modified Redfield Simulation.---}
The FMO complex is a PPC present in green sulfur bacteria which acts as a`molecular' wire by
mediating exciton transfer from the light-harvesting chlorosomes (antenna complex) to the
bacterial reaction center \cite{blankenship2002molecular,van2000photosynthetic}. It has a
trimeric structure with each of the three monomers containing seven chromophores bound to
a common protein structure. Under low light conditions, it can be assumed that there is only
one excitation at a time, and that all monomers operate independently. The excitation is
thought to originate at an eighth site (common to several monomers) which injects an exciton
locally at sites $1$ or $6$ of a given monomer \cite{eightsitepaper}. The exciton then relaxes
through the manifold of exciton states, and is directed towards site $3$ (which is the primary
component of the lowest energy exciton state). Due to its proximity to the reaction centre (RC),
the exciton on site $3$ is then rapidly transferred out of the FMO complex
\cite{blankenship2002molecular,van2000photosynthetic}. Experimental and theoretical studies have
described two main relaxation pathways in the FMO complex
\cite{brixner2005two,adolphsrenger06, ishizaki2009theoretical, chindc2010}. Here we shall focus
on the dominant pathway that proceeds through site $1$ to site $3$, although our numerical analysis is based on simulations
using the Modified Redfield approach as described in \cite{adolphsrenger06} of the entire eight-site FMO structure
with an initial excitation on site $8$. The sink is connected to site $3$. Unless stated otherwise,
the Hamiltonian parameters are taken from \cite{eightsitepaper}, and the spectral density of the
environment is based on Adolphs and Renger fit to fluorescence narrowing experimental data \cite{adolphsrenger06}.

\begin{figure}[h]
\begin{center}
\includegraphics[width=0.45\textwidth]{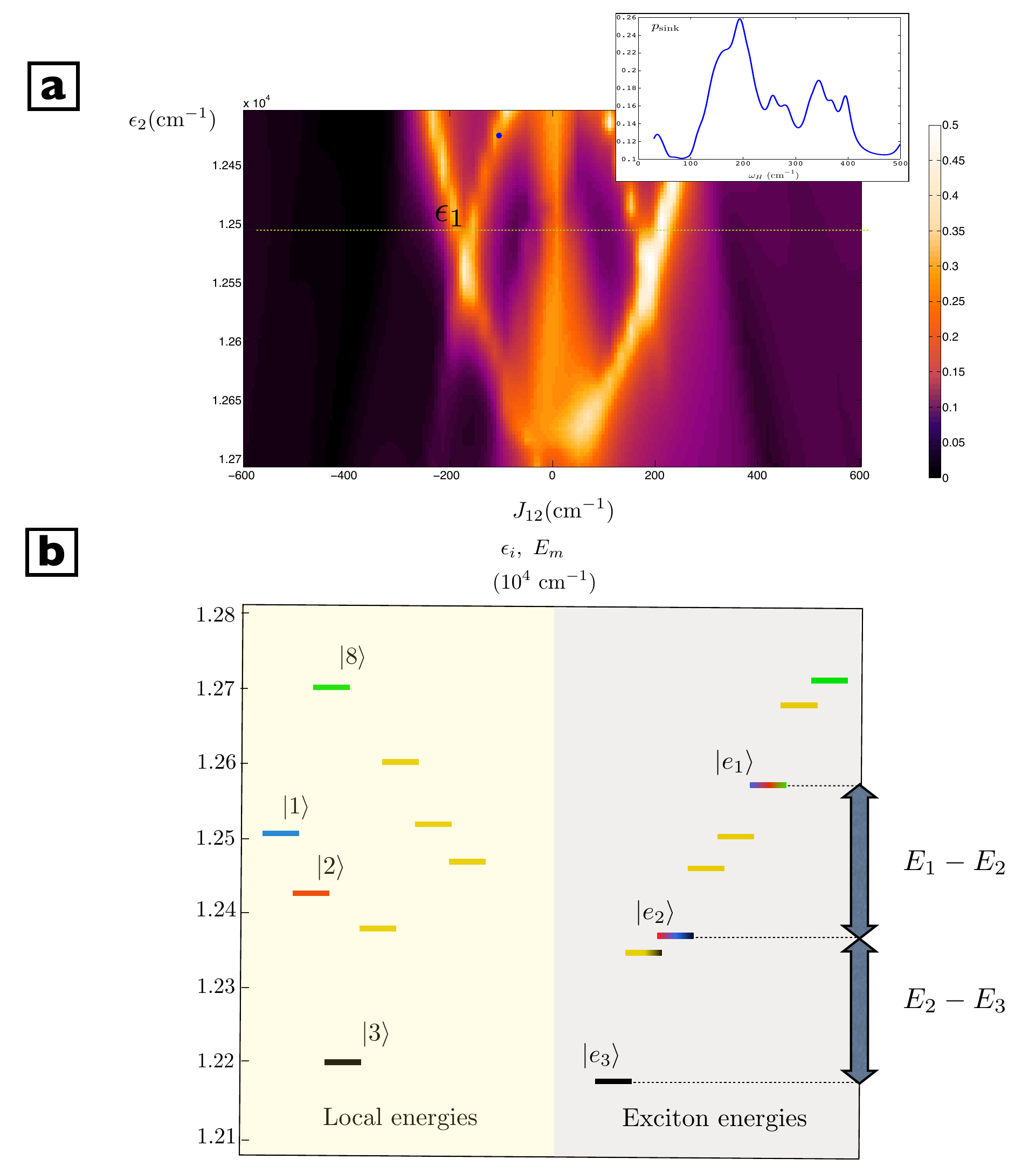}
\end{center}
\caption{a) Population of the sink, $p_{\text{sink}}(t=2 \text{ ps})$ for different values of the coupling between sites 1 and 2  ($J_{12}$) and the energy of site 2 ($\epsilon_2$). We clearly see that the actual values for $J_{12}$ and $\epsilon_2$ lie at the top of a maximum. The inset shows $p_{\text{sink}}(t=2 \text{ ps})$ for the actual couplings as observed for the FMO, assuming an spectral density function such as that of  \cite{supinfo} but peaked at $\omega_H$.  b) Local site and exciton energies for the $H_S$ values corresponding to the FMO, as taken from \cite{eightsitepaper}. The arrows show how the splitting might act as an antenna for the energy peak close to 180 $\text{ cm}^{-1}$}
\label{fig:nonmarkov8sitescouplings}
\end{figure}

Fig. \ref{fig:nonmarkov8sitescouplings}a shows the final population transferred to the sink
for a simulation of a total time $t=2\text{ ps}$ at $T=77\text{ K}$ as a function of $J_{12}$
and $\epsilon_2$. The transport efficiency shows a complex landscape of maxima and minima for
the parameter region explored, but we find that the Hamiltonian parameters of Ref. \cite{eightsitepaper}
(marked as a blue cross) are located at an optimal point. Figure \ref{fig:nonmarkov8sitescouplings}b
shows the energy level structure in both the local and exciton bases for the physical Hamiltonian
parameters of \cite{eightsitepaper}. We see in the exciton basis that the energy differences
corresponding to the relaxation transitions connecting sites $1, 2$ and $3$ are well matched
to the maximum of the experimentally-determined spectral function, and therefore this relaxation
channel forms a striking example of an operational phonon antenna in a biological transport
system. The inset of Fig.  \ref{fig:nonmarkov8sitescouplings} shows $p_\text{sink}$ for fixed
Hamiltonian parameters and a varying peak position $\omega_{H}$ in the spectral density (as
described in Appendix D). Again, we find that $p_{sink}$ is  a multi-peaked function, but the
most prominent maximum appears at $\approx 190 \mathrm{cm}^{-1}$, which is very close to the
experimentally-fitted peak in the spectral function of Adolphs and Renger ($180 \mathrm{cm}^{-1}$)
\cite{adolphsrenger06}.

\textit{The phonon antennae in a broader context ---} The phonon antennae principle which states
that \textit{electronic interactions are tuned to achieve resonance with maxima of the environmental
vibrational spectra and thus strongest response} may be viewed in a broader context in which
the tuning of electronic degrees of freedom to match the frequency of external signals is exploited
to achieve enhanced system response to facilitate sensing and transport.

In fact, the phonon antennae principle is closely related to the Hartmann-Hahn condition \cite{nmr}
which is of considerable importance in sensing devices. Here the challenge is that the sensor, e.g.
an electron spin, may not be resonant with an external system generating a signal (e.g. a different
electron or nuclear spin). This problem may be overcome by continuous driving of the electron spin in
the sensor at a strength that leads to a splitting of dressed states that matches the externally generated
signal and thus permits a strong response of the sensor (transitions between upper and lower dressed
state) \cite{London2012,NatPhys2012,NJP2012}.

A further example in which tuning is not achieved by driving is the recently proposed theory that olfaction
(our sense of smell) is explained, at least partially, as originating from phonon assisted electron transport
where electron source and drain in the receptor are separated by an energy splitting that is matched to a
specific vibrational frequency of an odorant molecule. Electron tunneling, hence strong response, will occur
only if the molecular vibrations matches this energy gap \cite{Turin1996}.

\textit{Discussion.---} Taken together, these results suggest that the physically important relaxation pathway between sites $1$ and $3$ is mediated by pigments which are spectrally and spatially positioned by the protein to efficiently sample the spectral function of the protein's fluctuations. Whether this optimality is a determinant in the emergence of this structure in nature is beyond the scope of this study, but it is striking how well the phonon antenna concept can be used to rationalise the site energies and couplings of the pigments participating in this pathway. Other considerations, such as structural stability and steric constraints imposed by the need to integrate the structure into the cellular environment are likely to drive the supramolecular structure of PPCs, however the multiplicity of local maxima we find \textit{does} provide the opportunity to find an optimal working point for transport in the \textit{vicinity} of a structure arising from these other factors. Indeed, we find that the FMO complex parameters correspond to a local maximum of the $1\rightarrow3$ pathways efficiency.  Interestingly, changes in temperature will also alter the inter-exciton energy that maximises the relaxation rates in Eq. (\ref{rate}), and an evolving or adaptive system might be able to change its coupling structure to maintain the inter-exciton energy differences at the maximum of the thermal noise strength.

We also point out that while the Redfield formalism allows us to give a physically clear description of the essential phonon antenna mechanism, the use of Markovian Redfield theory for PPC problems neglects a number of important dynamical effects (environmental memory, stokes shift, discrete modes in the spectral function) which are crucial for understanding the \textit{persistence} of inter-eigenstate coherence that is generated in experiments \cite{chin2012vibrational,kramer12, christensson2012origin}. These effects have been neglected here for simplicity, but exploring these effects in the context of the phonon antenna mechanism could lead to an even more sophisticated principle, involving transient, broad-band sampling of the spectral function and evolving exciton spectra and localisation lengths which could, potentially, activate or deactivate relaxation pathways dynamically. Finally, the concept of using coherent interactions to alter the sampling of a (fixed) environmental spectral function could be readily achieved in a variety of atomic and condensed matter systems \cite{examples,London2012}, and could also be used as a sensitive probe for measuring spectral functions by measuring transport rates as functions of a controllable coupling parameter.

\textit{Acknowledgements.---}
This work was supported by the Alexander von Humboldt-Foundation, the EU STREP project PICC and the EU Integrated Project Q-ESSENCE. AWC acknowledges support from the Winton Programme for the Physics of Sustainability. M. del Rey was supported by the Spanish MICINN Project FIS2011-29287, CAM research consortium QUITEMAD S2009-ESP-1594, CSIC JAE-PREDOC2010 grant and the Fundaci\'on Bot\'in. The authors would also like to thank Rienk van Grondelle for fruitful discussions.

\newpage
\section{Supplementary information}

\subsection{Appendix A: Detailed Model}

The total Hamiltonian ($\hbar=1$) describing electron-phonon interactions consists of the \textit{system} Hamiltonian $H_S$ which includes the electronic degrees of freedom, the free Hamiltonian $H_R$ of the phonon bath describing fluctuations in the protein structure, considered as a set of $N$ independent bosonic baths $\{i=1,2...N\}$ coupled to each of the sites (chromophores), and the linear coupling of fluctuations to the excitations $V$:

\begin{align}
H &\!=H_S+H_R+V,\ \  H_S = \sum_i \epsilon_i \ket{i}\!\!\bra{i} + \sum_{i\neq j } J_{ij} \ket{i}\!\!\bra{j} \nonumber\\
H_R &=\! \sum_i\! \sum_k  \omega_k a^\dagger_{ik} a_{ik}, \ \  V = \sum_{ik} g_{ik} (a_{ik}+a^\dagger_{ik}) \ket{i}\!\!\bra{i} \label{mainequation}
\end{align}
where $a_{ik}, a^\dagger_{ik}$ are the usual creation and annihilation operators, respectively, satisfying $[a_{ik}, a^\dagger_{ik}] = \delta_{ik}$. For simplicity, we will assume an identical coupling to the environment for each site $g_{ik} = g_k$  with $ \mathcal J(\omega) = \sum_k g^2_{k} \delta(\omega - \omega_k)$ being the vibrational spectral density.

In  the exciton basis,  $H_S$ is diagonal,  $H_S\ket{e_n} = E_n \ket{e_n}$,  and we will write $\ket{i}=\sum_n C_n^i \ket{e_n}$.
In that basis the Hamiltonian can be expressed as $H_S = \sum_n E_n \ket{e_n}\bra{e_n}$ and $V = \sum_{m\neq n} \Big[ \sum_{ik}  g_{ik} C^{i}_m C^{i^*}_n (a_{ik}+a^\dagger_{ik}) \Big] \ket{e_m}\bra{e_n}$, where we have neglected the Lamb and Stark shifts corresponding  to $m=n$ in the last term.

A state $|0\rangle$, or sink, has been introduce to model the transfer to the RC, where the excitation will end up eventually as a relaxation process coming from site 3, so the behaviour of the system will account for an additional channel $\Gamma_{3 \rightarrow 0} = 1/{(1 \ ps)}$. Although not fully equivalent, when using the master equation in the exciton picture we will model this irreversible transfer to the sink as happening at rate  $\Gamma_{e_n \rightarrow 0} = |\braket{e_n}{3}|^2\Gamma_{3 \rightarrow 0} $.

In all cases we study the time evolution of the sink's population following the selective excitation of site $\alpha$, $\ket{\psi}_0 = \ket{\alpha}$. In the considered 3 sites model system, $\alpha=1$, while in the 8 site model of FMO, it is $\alpha=8$.

\subsection{Appendix B: Non-Markovian analysis of a three site network}

In this section we analyse the 3-site case beyond the concerns exposed in the markovian treatment, with the aim to show how manifestations of the phonon antenna also occur in a paradigmatic example of non-markovianity. For that matter we will assume that all levels in the network are strongly coupled to identical independent harmonic oscillators, all tuned to the same frequency $\omega_H$, representing therefore the coupling to a unique discrete mode. The coupling between sites would be given by the same matrix $H_S$.

\begin{figure}[h]
\begin{center}
\includegraphics[width=0.3\textwidth]{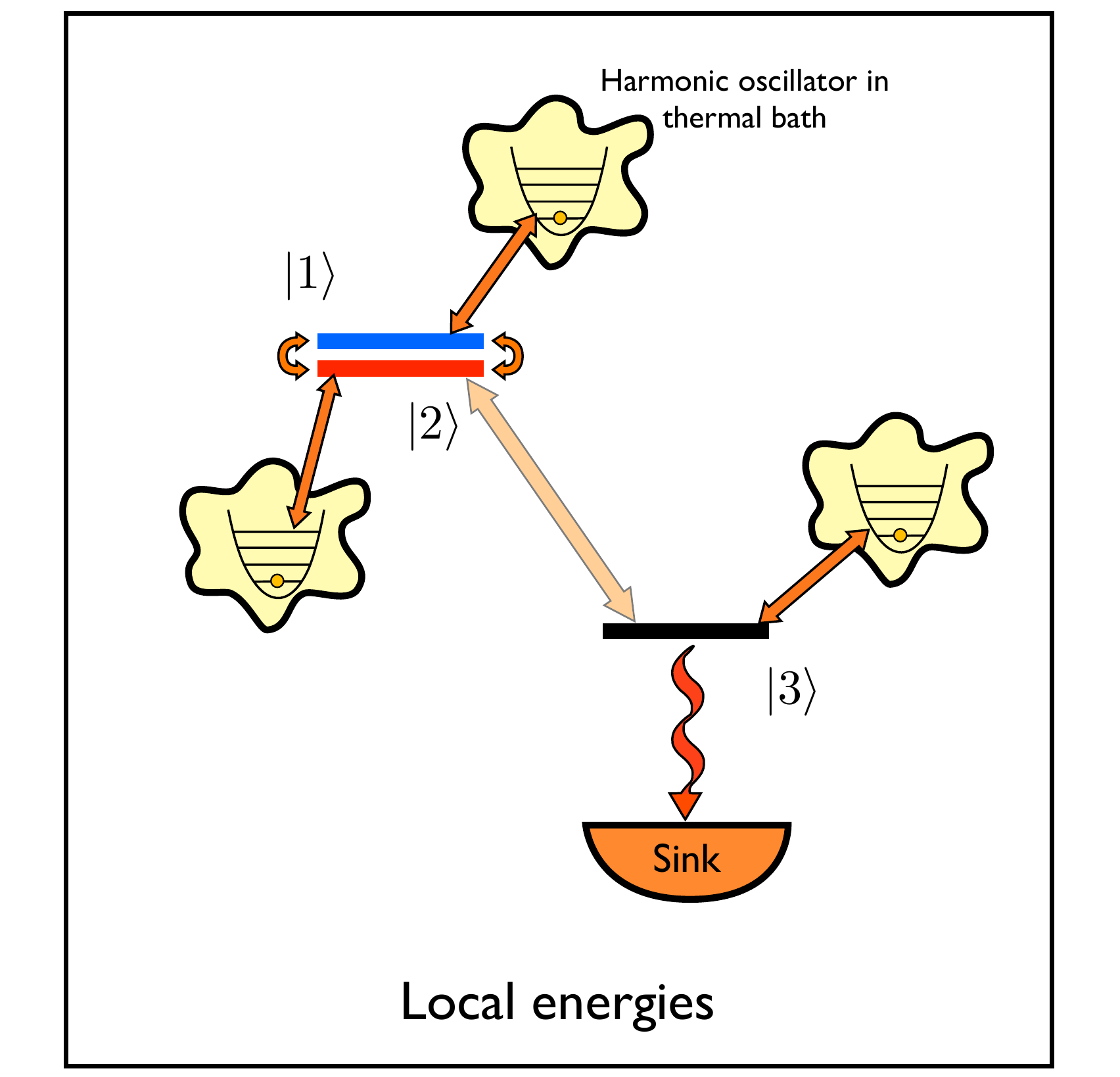}
\end{center}
\caption{Evaluation of non-Markovian effects. Local sites are coupled to damped harmonic oscillators, so that the reduced dynamics for the electronic degrees of freedom alone is in general non reproducible
with a Markovian master equation.}
\label{fig:antennaeffect2}
\end{figure}

This model presents the advantage of enabling a direct analysis in the local site basis, which lets us \textit{avoid} making the approximation $\Gamma_{e_n \rightarrow 0} = |\braket{e_n}{3}|^2\Gamma_{3 \rightarrow 0} $ for the transfer rate to the sink node. The local Hamiltonian described in Eq. (\ref{mainequation}) fits this model if we assume that the sum of $k$'s restricts to just one mode.
The dissipation of each mode will be modelled using a master equation formalism, where the harmonic oscillators relax at the same rate ($\Gamma=60\text{ cm}^{-1}$) to a (Markovian) thermal bath at temperature $T$, according to the Lindblad superoperator:

\begin{multline}
\mathcal{L}_{\text{diss}}=\sum_i \Gamma [n(\omega_H)+1] [a_i \rho a_i^\dagger - \frac{1}{2}a_i^\dagger  a_i \rho- \frac{1}{2} \rho a_i^\dagger  a_i] \\
+\sum_i \Gamma [n(\omega_H)] [a_i^\dagger \rho a_i  - \frac{1}{2} a_i a_i^\dagger \rho- \frac{1}{2} \rho  a_ia_i^\dagger ]
\end{multline}

The coupling constant $g$ is chosen to match the quasi Lorentzian profile previously presented in \cite{garg}. There, it is shown that a two level system coupled to a bath with that profile behaves equivalently as if it were coupled to a damped oscillator with coupling constant $g = \omega_H \sqrt{\frac{\beta}{8 \kappa}} \simeq \frac{1}{2} \sqrt{\lambda\omega_H} \sim 80 \text{ cm}^{-1} $
Within this approach, the oscillator is damped within an environment of Ohmic nature $\mathcal{J}(\omega) =\frac{\Gamma}{\omega_H}\omega$. Although that is not exactly our case, it can be used as a first approximation to study the impact of non-Markovian features.

The transfer to the sink is modelled as an irreversible process from site 3 to site 0 with $\Gamma_{3 \rightarrow 0} = 1/{(1 \ ps)}$ as described by:
\begin{equation}
 \mathcal{L}_{\text{sink}}=\Gamma_{3 \rightarrow 0}\Big[|0\rangle \langle 3| \rho |3\rangle \langle 0| - \frac{1}{2} |3\rangle \langle 3|\rho- \frac{1}{2} \rho|3\rangle \langle 3|\Big]
 \end{equation}

\begin{figure}[h]
\begin{center}
\includegraphics[width=0.45\textwidth]{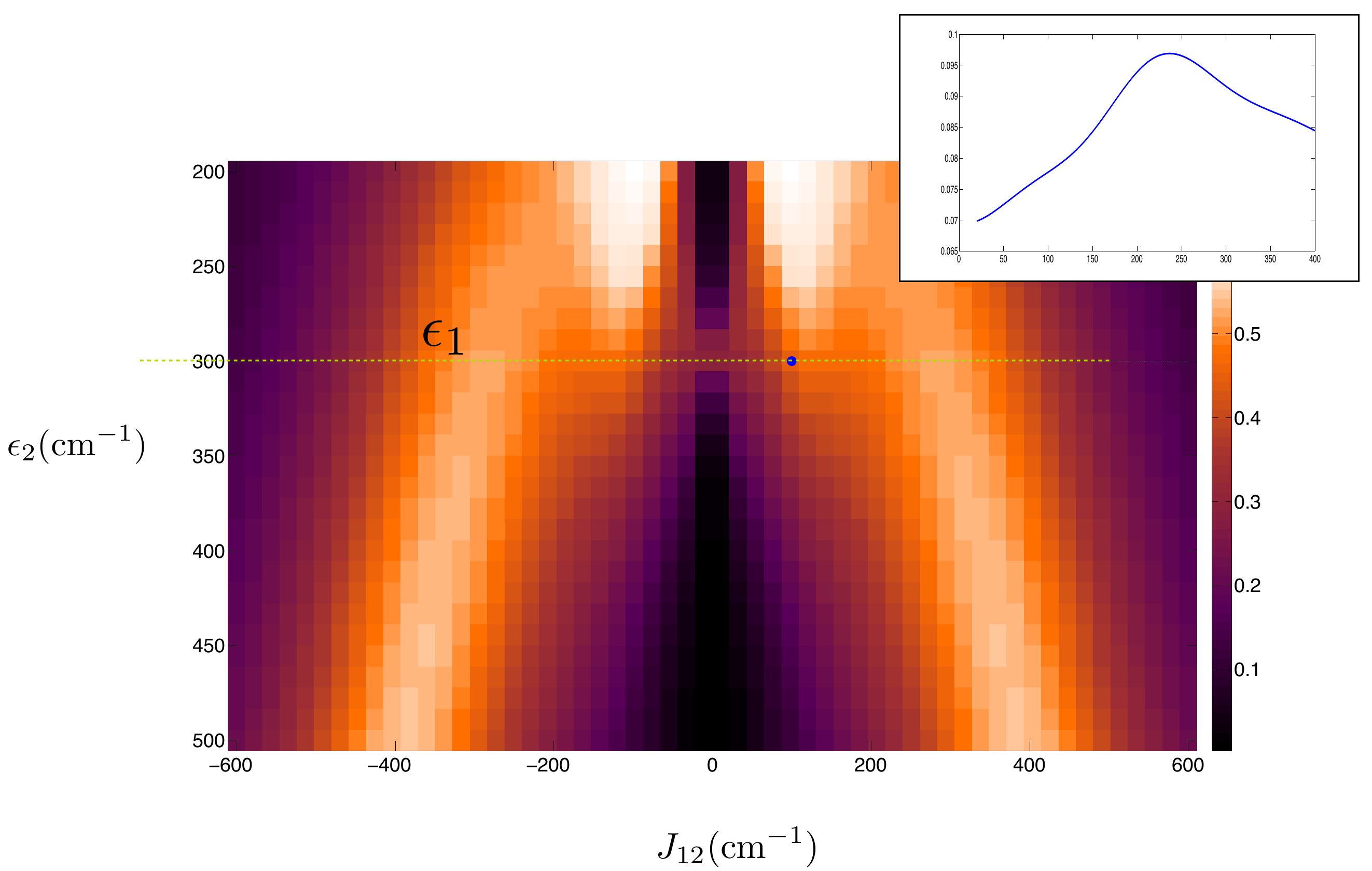}
\end{center}
\caption{Simulation of 10 ps transfer with oscillators tuned at 200 $\text{ cm}^{-1}$ at T=4 K. The actual couplings lie close to a maximum. The inset shows how a phonon antenna effect appears if the mode is tuned at roughly 237 $\text{ cm}^{-1}$.}
\label{fig:nonmarkov3sitescouplings}
\end{figure}

The inset of Fig. \ref{fig:nonmarkov3sitescouplings} shows some changes in the phenomenon. It takes place close to the expected value, although the shift is prominent. The explanation relies on the direct coupling to the harmonic oscillators, which is so strong that leads to a modification in the excitonic levels of the whole system, the final configuration being way more complicated than the one explained in section II. The concept remains valid, but of course the intuition about the configuration of the antenna is not so precise.

Fig. \ref{fig:nonmarkov3sitescouplings} shows a plot with the mode fixed at 200 $\text{ cm}^{-1}$. As we see, the values of the Hamiltonian are close to a maximum in transfer yield showing once more the forming of an antenna.

\subsection{Appendix C: Spectral function}

For the sake of modelling appropriately the spectral function used for the FMO case, we have relied on the function proposed by Adolphs and Renger \cite{adolphsrenger06}. In that reference a combination of super-ohmic densities is used and then a single effective high-energy vibrational mode $H$ is added with Huang Rhys factor $S_H=0.22$ and $\omega_H=180 \text{ cm}^{-1}$.
In our case, instead of using a delta function, we introduce a pure Lorentzian with the same characteristics but a linewidth of $\Gamma_H=60 \text{ cm}^{-1}$.
With those prescriptions the spectral function can be written as

\begin{multline}
\mathcal J (\omega) = \frac{\lambda[1000\omega^5 e^{(-\omega/\omega_1)^{-1/2}} + 4.3\omega^5 e^{(-\omega/\omega_2)^{-1/2}}]}{9!(1000\omega_1^5+4.3\omega_2^5)} \\ +\frac{ S_H \omega_H^2}{\pi} \frac{\Gamma_H}{(\omega-\omega_H)^2+\Gamma_H^2}, \label{densityfunction}
\end{multline}

where $\lambda=35 \text{ cm}^{-1}$ is the reorganization energy for the background, $\omega_1=0.5 \text{ cm}^{-1}$,  $\omega_2=1.95 \text{ cm}^{-1}$ and $\omega_H=180 \text{ cm}^{-1}$

\subsection{Appendix D: Quantitative Figure of merit}
\begin{figure}[h]
\begin{center}
\includegraphics[width=0.45\textwidth]{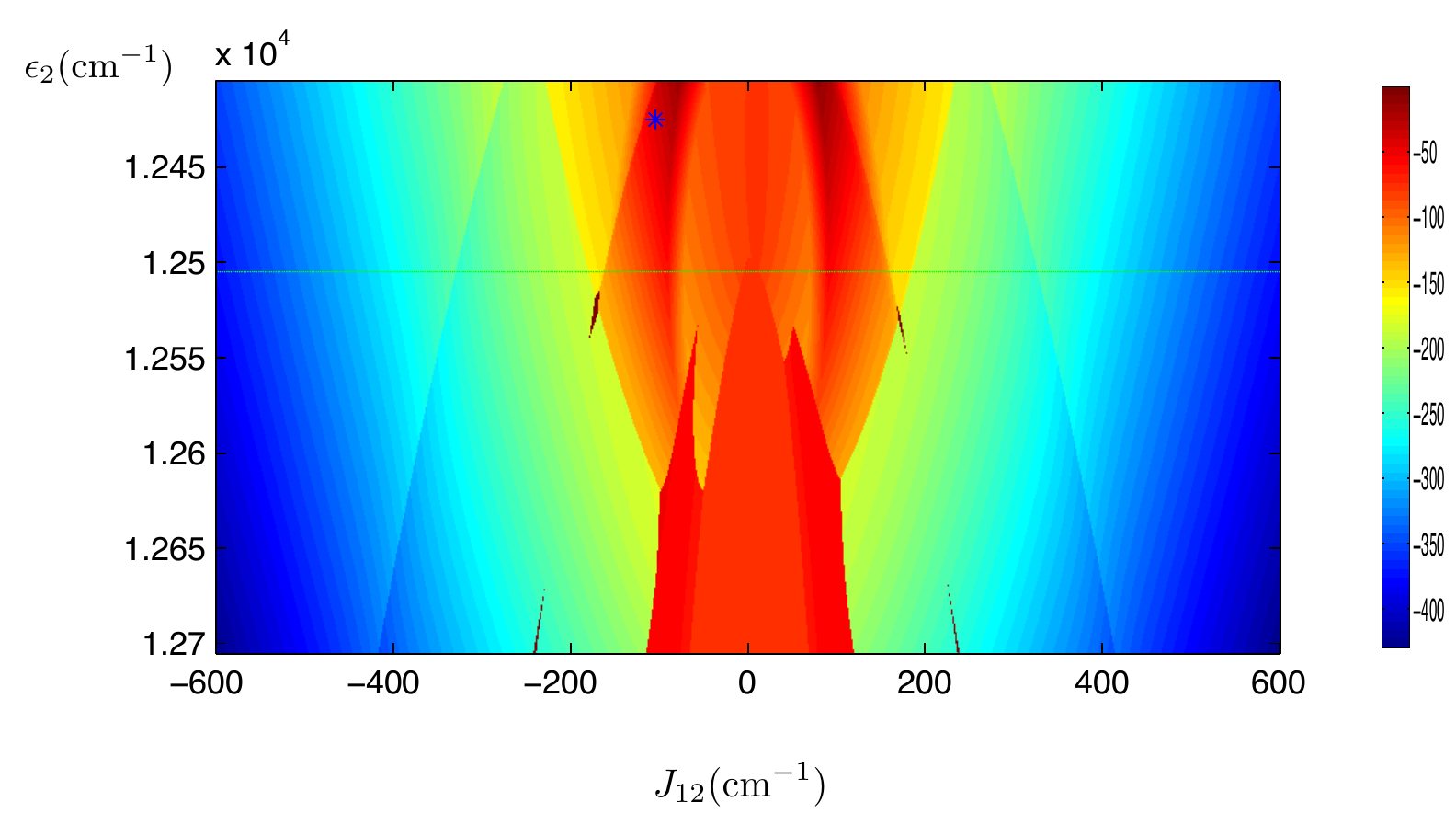}
\end{center}
\caption{$F_{\text{antenna}}$ for varying values of $J_{12}$ and $\epsilon_2$ with $\omega_H= 180 \text{ cm}^{-1}$ as in Fig. \ref{fig:nonmarkov8sitescouplings}a}
\label{fig:figureofmerit}
\end{figure}

We have envisioned a figure of merit to check how well can the concept of  \textit{phonon antenna }alone predict the main features of the landscape as shown in Fig. \ref{fig:nonmarkov8sitescouplings}a. For the same variations of $J_{12}$ and $\epsilon_{2}$ we have studied  the level structure and computed the simple value:
\begin{multline}
F_{\text{antenna}}=\text{max \{} -|\omega_H -|E_+-E_-||-|\omega_H -|E_--E_G||, \\-\frac{1}{2}|2\omega_H -|E_+-E_G|| \}
\end{multline}
The higher the value of $F_\text{antenna}$, the more assisted by the  environment the transition to the sink would be, as inferred from the scheme described.  Fig. \ref{fig:figureofmerit} shows the map of $F_\text{antenna}$ in the same range of parameter values as Fig. \ref{fig:nonmarkov8sitescouplings}.

\end{document}